\documentstyle{article}

\def\nn{\nonumber\\[3pt]}
\def\eq{\begin{equation}} \def\qe{\end{equation}}
\def\Ker{\mbox{\rm Ker}\,} \def\Aut{\mbox{\rm Aut}\,}
\def\Hom{\mbox{\rm Hom}} 
\def\eqa{\begin{eqnarray}} \def\aqe{\end{eqnarray}}
\def\eqas{\begin{eqnarray*}} \def\saqe{\end{eqnarray*}}
\def\arr{\begin{array}} \def\rra{\end{array}}
\def\hl{\hline} \def\hh{\hl\hl}
\def\nlh{\\\hl} \def\nlhh{\\\hh}
\def\nlsh{\\[7.5pt]\hl} \def\nlshh{\\[7.5pt]\hh}

\def\mathz{\mbox{\sf Z\hspace{-0.90ex}Z}}
\def\mathc{\mbox{\sf C\hspace{-1.00ex}l\hspace{0.45ex}}}
\def\ve{\varepsilon}
\def\al{\alpha} \def\be{\beta} \def\ga{\gamma}
\def\om{\omega} 
\def\uo{U(1)}
\def\zn{\mathz_N} \def\zz{\mathz_2}  \def\zv{\mathz_4}
\def\zi#1{\mathz_{N_{#1}}} \def\zni{\zi{i}}
\def\bim{\bibitem}
\def\ie{{\em ie}} \def\cf{{\em cf}} \def\viz{{\em viz}}
  \def\eg{{\em eg}} \def\cf{{\em cf}}

\def\ii{\hat{\imath}}
\def\cp{{\cal P}} \def\cq{{\cal Q}}
\def\ket#1{\left|#1\right>} 
\def\bct#1#2{\left<#1|#2\right>}
\def\um{-\!} 
\def\px{\pi_{cx}} \def\py{\pi_{cy}}

\def\ru{\rule{0pt}{16pt}}

\begin{document}
\newfont{\abstr}{cmr9}
\thispagestyle{plain}
\begin{center}
{\large\bf Magnetic translation group as group extensions}\\ [6pt]
{\sc Wojciech Florek} \\ [4pt]
A. Mickiewicz University, Institute of Physics\\
\end{center}


\hfill\begin{minipage}{11.5cm}{{\abstr 
Extensions of a direct product $T$ of two cyclic groups $\zi1$ and $\zi2$ by
an Abelian (gauge) group $G$ with the trivial action of $T$ on $G$ are
considered. All possible (nonequivalent) factor systems are determined using
the Mac Lane method. Some of resulting groups describe magnetic translation
groups. As examples extensions with $G=\uo$ and $G=\zn$ are considered and
discussed. }} \end{minipage}

\section{Introduction}
The idea of magnetic translation groups, appearing in considerations of
movement of electron in an external magnetic field, was proposed 
independently by Brown \cite{bro} and Zak \cite{zak}. From those works
follows that the magnetic translation group is an image of Weyl--Heisenberg
group \cite{wey} obtained by imposing the Born--von K\'arm\'an periodic
boundary conditions. The general description of similar problems has been
presented by Schwinger \cite{sch}, who considered {\em unitary operator
bases}. One of the considered cases can be interpreted as a description of
{\em finite phase space}. This two-dimensional space is spanned by one space
(positional) dimension and the other corresponding to kinetic momentum. Two
unitary translation operators acting in these two dimensions are given by
exponential functions of Hermitian operators of momentum and position,
respectively, and, of course, they do not commute.  Algebraic structure
generated by such operators resembles an extension of a direct product of
two Abelian (translation) groups by a group $G$ containing commutators (or
simply factors since, in general, it is a subgroup of the field of complex
numbers). Therefore, such extensions are throughout studied in this work and
the physical relevance is indicated.

In the next section some basic ideas of the Weyl--Heisenberg group, finite
phase spaces and magnetic translations groups are briefly presented. All
possible (central) extensions of a direct product $T$ of two finite cyclic
groups $\zi1$ and $\zi2$ by an Abelian group $G$ are determined in Sec 3.
The Mac Lane method \cite{mac1,mac2} (see also Lulek \cite{lul,lch1,lch2}),
has been applied and the solution can be given in a general (analytic) form
(see also the appendix).  In Sec 4 the cases of $G=\uo$ and $G=\zn$ are
discussed as the simplest, and the most important, examples of possible
groups of factors (gauge groups).  This work is ended by a short discussion
and remarks.

\section{Basic Ideas}
\subsection{Weyl--Heisenberg Group} 
Let $Q$ and $P$ be two Hermitian canonically conjugated operators, \ie\ a
{\em complementary} pair of operators, \viz
\eq [Q,P] = \ii\hbar. \label{comm} \qe
It is natural to transfer this property to the unitary operators, which are
more accessible than the Hermitian ones. As a rule one constructs (unitary)
operators using the exponential function
\eq   \cq = \exp(\ii Q\al)\qquad{\rm and}\qquad \cp = \exp(\ii P\be),
\label{h2u} \qe
where $\al$ and $\be$ are real numbers (parameters). A group generated by
these operators is non-Abelian one since from the above formulae one
immediately obtains \cite{wey,thi}
\eq   \cp\cq=\cq\cp\exp(\ii\al\be\hbar). \label{weyrel} \qe
The Born--von K\'arm\'an periodic conditions (the same period $N$ for both
operators) yield a finite group generated by two operators $U$ and $V$ with
the following relations
\eq   UV=VU\ve,\qquad U^N=V^N=E,\qquad \ve^N=1, \label{finrel}  \qe
so $\ve$ is the $N$-th root of $1\in\mathc^{*}$. This group, roughly
speaking, is a {\em magnetic translation group} (\cf\/ \cite{wal}). The
relation (\ref{weyrel}) can be considered as a basis for the Weyl algebra
\cite{wey,thi} and its finite counter-part (\ref{finrel}) has been
investigated by Schwinger \cite{sch}.

\subsection{Finite Phase Space} \label{secsch}
Let us summarize the most important (in our considerations) results of
Schwinger's work \cite{sch}. In a finite-dimensional eigenspace of a given
Hermitian operator (say $P$) with a basis $\{\ket{i}\mid 0\leq i<N\}$ a
unitary operator of the cyclic permutation can be introduced, \ie
\eq     V\ket{i} = \ket{i+1\,\bmod\, N}, \qe
so, in general,
\eq   V^n\ket{i} = \ket{i+n\,\bmod\, N}  \qe
and
\eq   V^N\ket{i} = \ket{i}, \quad \mbox{\rm so}\quad V^N=E  \qe
($E$ is the identity operator in the $N$-dimensional space). The eigenvalues
of $V$ obey the same equation, \ie\/ $v^N_k=1$, and they are given by the
$N$ distinct complex numbers
\eq   v_k=\ve^k,\qquad k=0,1,\ldots,N-1,  \qe
where $\ve=\exp(2\pi\ii/N)$. The corresponding eigenvectors are given as the
following linear combinations 
\eq    \ket{k}_V=\frac{1}{\sqrt{N}}\sum_{i=0}^{N-1}\ve^{\um ki}\ket{i},
\label{Vv}  \qe
so $V\ket{k}_V=v_k\ket{k}_V$, and this equation is simply a finite version
of the Fourier transformation. In the new basis $\{\ket{k}_V\mid 0\leq
k<N\}$ one can define another unitary operator $U$ by the (`anty'-)cyclic
permutation, \ie
\eq   U\ket{k}_V=\ket{k-1\,\bmod\,N}_V.  \qe
The properties of this operator are the same as for $V$, but now the
eigenvectors are given as 
\eq    \ket{l}_U=\frac{1}{\sqrt{N}}\sum_{k=0}^{N-1}\ve^{lk}\ket{k}_V.
\label{Uv}   \qe
Combining (\ref{Vv}) and (\ref{Uv}) one obtains
\eq  \ket{l}_U=\frac{1}{N}\sum_{k=0}^{N-1}\ve^{k(i-l)}\ket{i}=\ket{l}, \qe
what can be easily presumed since (\ref{Uv}) describes the inverse of the
(finite) Fourier transformation (\ref{Vv}). The considered system of
(unitary) operators and their eigenvectors fulfil the following conditions
\eq    \arr[b]{rclcrcl} \rule[-3pt]{0pt}{\baselineskip}_V\!\bct{k}{l}_U & = 
 & \frac{1}{\sqrt{N}}\ve^{kl},&\qquad
 & \rule[-3pt]{0pt}{\baselineskip}_U\!\bct{l}{k}_V & = 
 & \frac{1}{\sqrt{N}}\ve^{-kl},\\[5mm] V^N&=&E,&&U^N&=&E,\\[5mm]
 UV&=&VU\ve,&&U^lV^k&=&V^kU^l\ve^{kl}.
\rra \label{epsi} \qe
Some properties of such a system and the physical relevance are discussed in
more details by Schwinger \cite{sch}. For our aim it is important that it
can described as an extension of groups. It should be stressed that though
$\ve\in\uo\subset\mathc^{*}$ we can limit ourselves to the cyclic
(multiplicative) group $C_N\simeq\zn$ generated by the primitive $N$-th root
of $1\in\mathc^{*}$.

\subsection{Magnetic Translation Group}
In the case of 2-dimensional magnetic translation groups the roles of the
Hermitian operators $Q$ and $P$ are played by the operators $\px$ and $\py$,
which are connected with coordinates of the center of the magnetic orbit,
\ie\ the orbit of an electron in a magnetic field (see \cite{wal} and
\cite{daza} for details). Strictly speaking, there are the following
operators ($\vec{H}$ is a uniform magnetic field along the $z$ axis):

\eqa 
\vec{\pi} &=& \vec{p}+\frac{e}{2c}(\vec{H}\times\vec{r}), \qquad 
 \mbox{\rm kinetic momentum}, \\[5pt]
\vec{\pi}_c &=& \vec{p}-\frac{e}{2c}(\vec{H}\times\vec{r}), \qquad 
 \mbox{\rm center of magnetic orbit}, 
\aqe \eq \left.\arr{lcr}
T_x(a)&=&\exp\left(\frac{\ii}{\hbar}\px a\right)\\[5pt]
T_y(b)&=&\exp\left(\frac{\ii}{\hbar}\py b\right)
\rra\right\}\qquad\mbox{\rm magnetic translations}
\label{magt} \qe
and
\eqa  Q = &y_0& = \px\frac{c}{eH},\\[5pt]
      P = &-{\displaystyle \frac{eH}{c}}x_0& =\py, \aqe
where the pair $(x_0,y_0)$ gives the coordinates of the center of the
magnetic orbit (see \cite{daza} and references quoted therein). The
parameters $a$ and $b$ determine the exponential transformation (\ref{magt})
and, after imposing the Born--von K\'arm\'an periodic conditions, correspond
to lattice constants. The last pair of Hermitian operators preserves the
commutation relations between the position and momentum coordinates, \ie\
\eq    QP - PQ = \ii\hbar   \qe
and their images under transformation (\ref{h2u}) are the magnetic
translations (\ref{magt}), \viz
\eqa  \cq = \exp(\ii Q\al) &=& T_x\left(\frac{\al\hbar c}{eH}\right) ,\nn
      \cp = \exp(\ii P\be) &=& T_y(\be\hbar). \aqe
 
\section{Group Extensions} 
The above presented brief summary of the most important results for the
magnetic translation group and a pair of complementary operators indicates
that magnetic translation groups can be described as an extension of a
direct product of two cyclic groups (of order $N$ and generated by $\cq$ and
$\cp$, respectively) by a `factor' (or `gauge') group $G$ being a subgroup
of the multiplicative group $\mathc^{*}$. It should be stressed that the
term {\em translation group} is a little misleading, since the unitary
operators $T_x$ and $T_y$ do not commute and they rather correspond to a
two-dimensional phase space with one direction connected with a position and
the second one with a momentum (so it is a pair of a one-dimensional space
$L$ and its adjoint $L^{*}$ rather than a product $L\otimes L$). However,
the algebraic structure of the translation group does not depend on it
because $L\cong L^{*}$.  Therefore in this section we investigate a general
problem of finding all non-equivalent extensions $G\Box(\zi1\otimes\zi2)$.
We assume $G$ to be an Abelian group due to the physical relevance
($\ve\in\mathc^{*}$) on the one hand, and, on the other hand, due to some
mathematical problems connected with non-Abelian extensions. The physical
applications suggest that the trivial action of $T$ on $G$ should be
considered, so --- strictly speaking --- a central extension
$G\bigcirc(\zi1\otimes\zi2)$ is investigated. The problem has been solved
applying the Mac~Lane method \cite{mac1,mac2} described also by Lulek 
\cite{lul} (for more details see the review articles \cite{lch1,lch2} and
references quoted therein).

\subsection{Alphabets and the Schreier Set} \label{trgr}
A two-dimensional finite translation group $T$ will be hereafter consider as
a direct product of two cyclic groups $\zni$, $N_i>1$, $i=1,2$. Therefore,
\eq   T=\left\{t=(t_1,t_2)\mid t_i\in\zni,\;i=1,2\right\}. \label{T}\qe
As a set of generators one can choose pairs
\eq  A\equiv\left\{\tau_1:=(1,0),\tau_2:=(0,1)\right\}. \label{tau}  \qe
In the Mac Lane method the second cohomology group $H^2(T,G)$, describing
all non-equivalent extensions of $T$ by $G$, can be found after
considerations of a free group $F$ and its normal subgroup $R$ --- a kernel
of the homomorphism $M:F\to T$. Moreover, one has to study the so-called
operator homomorphisms $\phi:R\to G$ and crossed homomorphisms $\ga:F\to G$.
In order to do it the alphabets $X$ (of $F$) and $Y$ (of $R$) have to be
found.

Let $F$ be a free group such that there exists a homomorphism $M:F\to T$ and
$M(X)=A$, where $X$ is an alphabet of $F$. Of course, $X$ consists of two
letters, say $x_1$ and $x_2$, with $M(x_i)=\tau_i$, $i=1,2$. For any
$n$-letter word $F\ni f= \prod_{i=1}^n \al_i^{\ve_i}$, where $\al_i\in X$
and $\ve_i=\pm1$, one obtains
\eq M(f)=\left((\sum_{i\in E_1}\ve_i)\bmod N_1,(\sum_{i\in E_2}\ve_i)
    \bmod N_2\right). \label{hoM} \qe
Subsets $E_j$ consist of indices $1\leq i\leq n$ such that $\al_i=x_j$ for
$j=1,2$. {\em Eg}, for $N_1=5$, $N_2=6$
$$  M(x_1^2x_2^{-1}x_1^{-1}x_2^4x_1^7x_2^{-5}x_1)=
    (9\bmod 5,-2\bmod 6)=(4,4). $$
The kernel $\Ker M \lhd F$, denoted hereafter as $R$, corresponds to group
relations imposed on generators $x_1$ and $x_2$. As representatives of {\em
right} cosets $R\backslash F$ the following elements are chosen 
\eq  f_{(t_1,t_2)} = \Psi(t_1,t_2) := x_1^{t_1}x_2^{t_2},  \label{Psi} \qe
where $\Psi:T\to F$ is such a mapping that $M\circ\Psi={\rm id}_T$. This
mapping determines also a choice function
\eq   \be := \Psi\circ M,\qquad \be:F\to F,  \label{beta}  \qe
which maps each element $f\in F$ onto the corresponding (right-)coset
representative $f_t$, where $t=M(f)$. These elements form the so-called
Schreier set
\eq    S := \{x_1^{t_1}x_2^{t_2}\mid t_i\in\zni, i=1,2\}.  \label{S} \qe
In general, the coset representatives do not form a group, but 
\eq  f_tf_{t'}=\rho(t,t')f_{tt'}, \label{rho} \qe
where $\rho(t,t')\in R$. These elements determine a factor system
$m:T\otimes T\to G$, then a group extension $G\bigcirc T$ {\em via} an
operator homomorphism (see below).

The alphabet $Y$ of the kernel $R$ can be chosen as nontrivial different 
factors $\rho(t,t')= f_tf_{t'}f_{tt'}^{-1}$ for $t\in T$ and $t'\in A$.
According to the Nielsen--Schreier theorem there are 
\eq   |Y|=1+(|X|-1)|T|  \qe
letters in this alphabet, so (in the considered case) one obtains
$|Y|=N_1N_2+1$. It is straightforward matter to show that these letters are
given by the following formulae
\eq  \arr[b]{rclcl}
A_{t_2} &=& x_1^{N_1-1}x_2^{t_2}x_1x_2^{-t_2} &\;\;{\rm for}\;\;&
 0\leq t_2<N_2,\\[5mm]
B_{t_1} &=& x_1^{t_1}x_2^{N_2}x_1^{-t_1} &{\rm for}& 0\leq t_1<N_1,\\[5mm]
C_{t_1t_2} &=& x_1^{t_1}x_2^{t_2}x_1x_2^{-t_2}x_1^{-t_1-1}& {\rm for}&
 0\leq t_1<N_1-1\;\;{\rm and}\;\;1\leq t_2<N_2. \rra\label{AF}  \qe

All factors $\rho(t,t')$ can be written in this alphabet since they are
elements of the kernel $R$. Let $f\in R$ be a word given in the alphabet
$X$. To find out its `spelling' in the alphabet $Y$ one can use the
`translation' formula
\eq  f= \prod_{i=1}^n \al_i^{\ve_i}= \prod_{i=1}^n 
 \be(f_{i-1})\al_i^{\ve_i}\be(f_{i-1}\al_i^{\ve_i})^{-1}, \label{trans} \qe
where $f_i$ denotes an initial subword of $f$ consisting of the first $i$ 
letters ($f_0:=1_F$). Each nontrivial factor in the above product is either
a letter of the alphabet $Y$ or the inverse of a letter (\ie\, an element of
$Y^{-1}$). Introducing a new set of letters  
\eq  D_{t_1t_2} := \prod_{i=0}^{t_1-1}C_{it_2},\qquad D_{0t_2}:=1_F, \qe
all factors $\rho(t,t')$ can be written as
\eq \rho\left((t_1,t_2),(t'_1,t'_2)\right) = D_{t_1t_2}^{-1}
   (D_{N_1t_2}A_0)^{e_1}D_{t''_1t_2}(D_{t''_1N_2}B_0)^{e_2}\label{rhoY}\qe
where  
\eq t''_i = t_i + t'_i \bmod N_i = t_i + t'_i - e_i N_i  \qe
and 
$$  e_i=\cases {1 & if $t_i+t'_i\geq N_i$, \cr 0 & otherwise.}  $$
Other properties of the letters $A$, $B$, $C$, and $D$ are gathered in the
appendix.

\subsection{Operator and Crossed Homomorphisms}
An operator homomorphism $\phi:R\to G$ fulfils the condition
\eq  \phi(frf^{-1})=(\Delta\circ M)(f)(\phi(r)),  \qe
where $\Delta:T\to \Aut G$ describes an action of $T$ on $G$. For the
trivial action ($\Delta(f)={\rm id}_G$) one obtains 
\eq \phi(frf^{-1})=\phi(r).  \label{ophom} \qe
Each homomorphism $\phi$ is determined by its values for $f=x\in X$ and it
is enough to consider $r=y\in Y$. The elements $xyx^{-1}$, $x\in X$, $y\in
Y$ are gathered in Table \ref{tact}. The set of equations (\ref{ophom}), 
solved in an Abelian group $G$, provides us with the following conditions
\eq  \arr[b]{rclcrcl}
a_{t_2} &=& a_0+c_{0t_2}, &\qquad& a_{t_2} &=& a_{t_2+1}+d_{N_1-1,1},\\[5pt]
b_{t_1} &=& b_{t_1+1}, && b_{t_1} &=& b_{t_1},\\[5pt]
c_{t_1t_2} &=& c_{t_1+1,t_2}, && c_{t_1t_2} &=& c_{t_1,t_2+1}-c_{t_11},
\rra \qe
where the lower-case letters denote images of the upper-case letters:
$a_{t_2}=\phi(A_{t_2})$, $b_{t_1}=\phi(B_{t_1})$,
$c_{t_1t_2}=\phi(C_{t_1t_2})$. The solution can be written as
\eq a_{t_2} = a + t_2c, \qquad b_{t_1} = b, \qquad c_{t_1t_2} = t_2c, 
\label{ct} \qe
so $d_{t_1t_2}=\phi(D_{t_1t_2})=t_1t_2c$. The parameters $a\equiv a_0$ and 
$b\equiv b_0$ are any elements of the factor group $G$, but the parameter
$c\in G$ fulfils the condition
\eq  N_1c = N_2c = 0. \label{econd} \qe
Therefore, non-trivial solutions for $c$ exist if and only if 
$\gcd(N_1,N_2)=M\neq1$ and there is an element $g\in G$ with order dividing
$M$. It means that nontrivial solutions of (\ref{econd}) are possible if and
only if $G$ has a torsion subgroup.

\begin{table} \caption[]{\label{tact} Action of $X$ on $Y$}
\begin{center} \begin{tabular}{||c||c|c||}\hh
\ru$y$ & $x_1yx_1^{-1}$ & $x_2yx_2^{-1}$ \nlshh
\ru$A_{t_2}$ & $A_0C_{0t_2}$ & $D_{N_1-1,1}A_{t_2+1}$ \nlsh
\ru$B_{t_1}$ & $B_{t_1+1}$ & $D_{t_11}B_{t_1}D_{t_11}^{-1}$ \nlsh
\ru$C_{t_1t_2}$ & $C_{t_1+1,t_2}$ & $D_{t_11}C_{t_1t_2+1}D_{t_1+1,1}^{-1}$ 
  \nlshh
\end{tabular} \end{center} \end{table}

A crossed homomorphism $\ga$ is determined as a mapping satisfying the
following condition
$$  \ga(f,f')=\ga(f)+(\Delta\circ M)(f)(\ga(f')).  $$
Therefore, in the considered case, the crossed homomorphisms $\ga:F\to G$ 
become `ordinary' ones and they are determined by their values $\xi_1,\xi_2$
for the letters $x_1,x_2\in X$, respectively. For letters in the alphabet 
$Y$ one immediately obtains that
\eq \ga(A_{t_2}) = N_1\xi_1,\qquad \ga(B_{t_1}) = N_2\xi_2,\qquad
   \ga(C_{t_1t_2}) = \ga(D_{t_1t_2}) = 0.\label{Cp} \qe
It is evident that for a torsion-free group $G$ the non-zero values $\xi_i$,
$i=1,2$, yield the non-zero values of $\ga(A_{t_2})$ and $\ga(B_{t_1})$, but 
when there are the elements with orders being divisors of $N_1$ or/and $N_2$
then they can give $\ga(A_{t_2})$ or/and $\ga(B_{t_1})$ equal to $0$. 

The second cohomology group $H^2(T,G)$ can be found as a quotient group of
the group of operator homomorphisms $\Hom_F(R,G)$ and the group of crossed
homomorphisms $Z^1(F,G)$ restricted to $R$. Therefore, one has to find those
$\phi\in\Hom_F(R,G)$, which are not crossed homomorphisms.  The first
results is that nontrivial solutions of (\ref{econd}) lead to nontrivial
operator homomorphisms, since for all crossed homomorphisms one has
$\ga(C_{t_1t_2})=0$. However, you are remembered that it is possible for a
torsion or mixed group $G$ only. Moreover, the parameters $a$ and $b$ can
take any value, whereas $\ga(A_{t_2})$ and $\ga(B_{t_1})$ are limited by the
conditions (\ref{Cp}). These questions can be answered more precisely when
the arithmetic structure of numbers $N_1$ and $N_2$ are determined and the
gauge group $G$ is fixed. In the next section two, the most important,
examples are presented.

A factor system $m:T\times T\to G$ can be found as images of factors
$\rho(t,t')$ under nonequivalent operator homomorphisms $\phi$. Therefore, a
general formula for a factor system $m:T\times T\to G$ can be written as 
\eq m\left((t_1,t_2),(t'_1,t'_2)\right)=\phi\left(\rho((t_1,t_2),
  (t'_1,t'_2))\right)=e_1a + e_2b+ t'_1t_2c. \label{fs}  \qe
This formula is discussed in the next section for $G=\uo$ and $G=\zn$.

\section{Examples}
For the previous considerations follows that $\gcd(N_1,N_2)=M>1$, so we can
assume
\eq  N_1=k_1M,\qquad N_2=k_2M,\qquad\gcd(k_1,k_2)=1.  \label{nnm} \qe

\subsection{Unitary Group U(1)} 
Let 
\eq G=\uo=\left\{e^{\ii\delta}\mid0\leq\delta<2\pi\right\} \qe
(the multiplicative notion will be used hereafter what corresponds to the
addition of arguments $\delta$ modulo $2\pi$). 

When one assumes $c=1$ then all operator homomorphisms $\phi$, determined by
the parameters $a$ and $b$, are crossed ones, since equations
$a=\xi_1^{N_1}$ and $b=\xi_2^{N_2}$ can always be solved in $\uo$.  It
immediately follows that from these considerations that a number of 
nonequivalent extensions is equal to a number of solutions of the condition
(\ref{econd}). Then one has to find such $l_1$ and $l_2$ that
$$  c=\exp(2l_1\pi\ii/N_1)=\exp(2l_2\pi\ii/N_2). $$
Therefore, the following condition must be satisfied 
$$   \frac{l_1}{k_1}=\frac{l_2}{k_2};\qquad  0\leq l_i< Mk_i-1  $$ 
and one obtains 
$$  l_i=0,k_i,2k_i,\ldots,(M-1)k_i.  $$
As the final result all possible values of $\phi(C_{t_11})=c$ are found as
\eq c = \exp(2k\pi\ii/M)\quad {\rm for}\quad 0\leq k<M. \label{esol}\qe

It has been shown that there are $M=\gcd(N_1,N_2)$ nonequivalent extensions
$\uo\bigcirc(\zi1\otimes\zi2)$. As a representative of equivalent
factor systems this one with $a=b=1$ can be chosen. It follows from the
formula (\ref{fs}) that in this case the $k$-th factor system is given by
the following equation 
\eq m_k\left((t_1,t_2),(t'_1,t'_2)\right)= \exp(2\pi\ii kt'_1t_2/M) = 
  \om^{t'_1t_2k}, \label{u1fs}  \qe
where $\om=\exp(2\pi\ii/M)$ is the $M$-th root of 1. For any closed loop of
translations $(0,N_2-y),(N_1-x,0),(0,y)(x,0)$ ($x,y\neq0$) one immediately
obtains
\eqas &&[1,(0,N_2- y)][1,(N_1- x,0)][1,(0,y)][1,(x,0)]\\[5pt]&=&
 [\om^{xyk},(N_2- y,N_1- x)][\om^{xyk},(x,y)]\!=\![\om^{xyk},(0,0)]. \saqe
Therefore, a phase factor $\om^{xyk}$ corresponds to such a loop.  As an
example in Table \ref{ext4} the factor systems $m_1$ and $m_2$ for
$N_1=N_2=4$ are presented. 

\begin{table}[t] \caption{Factor systems for $\uo\bigcirc(\zv\otimes\zv)$;
columns (rows) are labelled by $t'_1$ ($t_2$) only, since values of
factors do not depend on $t'_2$ ($t_1$, respectively) \label{ext4}}
\begin{center} \begin{tabular}{||l||*{4}{r|}|}
\multicolumn{5}{c}{\small a) $k=1$}\\ \hh
\multicolumn{1}{||r||}{\raisebox{-5pt}{$t_2$}$t'_1$} & 0 & 1 & 2 & 3\nlhh
 0    & 1 &      1 &   1 &      1 \\
 1    & 1 &  $\ii$ & --1 & $-\ii$ \\
 2    & 1 &    --1 &   1 &    --1 \\
 3    & 1 & $-\ii$ & --1 &  $\ii$ \nlhh
\end{tabular} \hspace{3pc} \begin{tabular}{||l||*{4}{r|}|}
\multicolumn{5}{c}{\small b) $k=2$}\\ \hh
\multicolumn{1}{||r||}{\raisebox{-5pt}{$t_2$}$t'_1$} & 0 & 1 & 2 & 3\nlhh
 0    & 1 &   1 & 1 &   1 \\
 1    & 1 & --1 & 1 & --1 \\
 2    & 1 &   1 & 1 &   1 \\
 3    & 1 & --1 & 1 & --1 \nlhh
\end{tabular} \end{center} \end{table} 

\subsection{Finite Cyclic Groups}
Let $G=\zn$ and $M=\gcd(N_1,N_2)$ (as above). Introducing $M'=\gcd(N,M)$ all
integers $N,N_1,N_2$ can be written as
$$ N = kk'_1k'_2M',\qquad N_i = k_ik'_ik'M',\quad i=1,2, $$
where $M'=\gcd(N,N_1,N_2)$, $k'=\gcd(N_1,N_2)/M'$ and $k'_i=\gcd(N_i,N)/M'$.
It follows from the condition (\ref{econd}) that
\eq  c = lkk'_1k'_2\qquad{\rm with}\qquad 0\leq l<M'. \qe
Considering values of (crossed) homomorphisms one obtains that different
values of $\ga(A_{t_2})=N_1\xi_1$ and $\ga(B_{t_1})=N_2\xi_2$ are obtained
for $kk'_2$ and $kk'_1$ values of $\xi_1$ and $\xi_2$, respectively.
Therefore, nonequivalent extensions are determined by
\eq   0\leq a<M'k'_1,\qquad 0\leq b<M'k'_2. \qe
Hence, in the considered case there exist $(M')^3k'_1k'_2$ nonequivalent
extensions with factor systems given by (\ref{fs}). Some of them are
isomorphic, since when $a'=\al a$, $b'=\al b$ and $c'=\al c$, with
$\al\in\zn$, then a mapping
\eq   [t,(t_1,t_2)] \mapsto [\al t,(t_1,t_2)] \label{extiso} \qe
determines a group isomorphism (if multiplication by $\al$ is an
automorphism of $\zn$, \ie\ if and only if $\gcd(\al,N)=1$). Of course, 
other isomorphisms can also be found. 

It is evident that for all $c\neq0$ each closed loop with mixed translations
in the first and second directions gains an appropriate factor $xyc\bmod N$.
But in this case also loops along $x$-th or $y$-th direction gain a factor
connected with the other parameters $a$ and $b$, \viz
$$
N_1[0,(1,0)]=[0,(1,0)]+\ldots+[0,(1,0)]=[0,(N_1-1,0)]+[0,(1,0)]=[a,(0,0)].
$$

\begin{table}[t] \caption{\label{t444}Factor system for the extension
$\zv\bigcirc(\zv\otimes\zv)$ with $a=2$, $b=3$, and $c=1$; $t_1t_2=00$ 
and $t'_1t'_2=00$ are omitted}
\begin{center}  \begin{tabular}{||c||r|r|r*{3}{||r|r|r|r}||}\hl
\raisebox{-5pt}{$t_1t_2$}~$t'_1t'_2$ & 
10 & 20 & 30 & 01 & 11 & 21 & 31 & 02 & 12 & 22 & 32 & 03 & 13 & 23 & 33\nlh
10 & 0 & 0 & 2 &  0 & 0 & 0 & 2 &  0 & 0 & 0 & 2 &  0 & 0 & 0 & 2 \\
20 & 0 & 2 & 2 &  0 & 0 & 2 & 2 &  0 & 0 & 2 & 2 &  0 & 0 & 2 & 2 \\
30 & 2 & 2 & 2 &  0 & 2 & 2 & 2 &  0 & 2 & 2 & 2 &  0 & 2 & 2 & 2 \nlh                
                                               
01 & 1 & 2 & 3 & 0 & 1 & 2 & 3 & 0 & 1 & 2 & 3 & 3 & 0 & 1 & 2 \\
11 & 1 & 2 & 1 & 0 & 1 & 2 & 1 & 0 & 1 & 2 & 1 & 3 & 0 & 1 & 0 \\
21 & 1 & 0 & 1 & 0 & 1 & 0 & 1 & 0 & 1 & 0 & 1 & 3 & 0 & 3 & 0 \\
31 & 3 & 0 & 1 & 0 & 3 & 0 & 1 & 0 & 3 & 0 & 1 & 3 & 2 & 3 & 0 \nlh
02 & 2 & 0 & 2 & 0 & 2 & 0 & 2 & 3 & 1 & 3 & 1 & 3 & 1 & 3 & 1 \\
12 & 2 & 0 & 0 & 0 & 2 & 0 & 0 & 3 & 1 & 3 & 3 & 3 & 1 & 3 & 3 \\
22 & 2 & 2 & 0 & 0 & 2 & 2 & 0 & 3 & 1 & 1 & 3 & 3 & 1 & 1 & 3 \\
32 & 0 & 2 & 0 & 0 & 0 & 2 & 0 & 3 & 3 & 1 & 3 & 3 & 3 & 1 & 3 \nlh
03 & 3 & 2 & 1 & 3& 2 & 1 & 0 & 3& 2 & 1 & 0 & 3& 2 & 1 & 0 \\
13 & 3 & 2 & 3 & 3& 2 & 1 & 2 & 3& 2 & 1 & 2 & 3& 2 & 1 & 2 \\
23 & 3 & 0 & 3 & 3& 2 & 3 & 2 & 3& 2 & 3 & 2 & 3& 2 & 3 & 2 \\
33 & 1 & 0 & 3 & 3& 0 & 3 & 2 & 3& 0 & 3 & 2 & 3& 0 & 3 & 2 \nlh
\end{tabular}   \end{center}  \end{table}

From general properties of finite groups (especially abelian and cyclic
ones) follows that it is enough to consider a case when all numbers $N$,
$N_1$ and $N_2$ are powers of a prime integer $p$. In general, there are
possible three cases
\eq\arr{rrlrlrl} {\rm (I)} :& N=&p^{\al},& N_1=&p^{\al+\be+\ga},&
  N_2=&p^{\al+\ga};\\[5pt]
{\rm (II)} :& N=&p^{\al+\be},& N_1=&p^{\al+\be+\ga},&
N_2=&p^{\al};\\[5pt]
{\rm (III)} :& N=&p^{\al+\be+\ga},& N_1=&p^{\al+\be},&  N_2=&p^{\al}.
\rra\qe

In all cases $b$ and $c$ has a value from the set $\{0,1,\ldots,p^{\al}-1\}$
but $a=0,1,\ldots,p^{\al}-1$ only in the case (I). In the two other cases
$a=0,1,\ldots,p^{\al+\be}-1$. So, a number of nonequivalent extension is
$p^{3\al}$ in the first case and $p^{3\al+\be}$ in the second and third
cases. It should be underlined that these results do not depend on $\ga$.
The special case $\al=0$ yields a direct product in the first case and
extensions of two cyclic groups in the other cases. On the other hand, for
all cases (I), (II) and (III) the condition $\be=\ga=0$ gives the same type
of extensions, {\em viz} $\mathz_{p^{\al}}\bigcirc
(\mathz_{p^{\al}}\otimes\mathz_{p^{\al}})$ (number of nonequivalent
extension is, of course, equal to $p^{3\al}$). As an example the case 
$p^{\al}=4$ for $a=2$, $b=3$, and $c=1$ is presented in Table \ref{t444}.

\subsection{Classification of extensions}

The parameters $a$, $b$, and $c$ provide a classification scheme of all
nonequivalent extensions. The most rough way is to distinguish zero and
non-zero values of these parameters, what yields eight types of extensions.
Due to possible isomorphisms a number of completely different (\ie\/
non-isomorphic) extensions is less than 8. It can be shown considering the
simplest example $\zz\bigcirc(\zz\times\zz)\simeq\zz\bigcirc D_2$, which has
been studied in \cite{lch2}. In the presented considerations the parameter
$c$ plays a special role, since:
\begin{itemize}
\item It leads to noncommutativity of group elements in the extension,
  whereas for $c=0$ the obtained groups are Abelian ones;
\item The two other parameters correspond to full loops along $x$ and,
  respectively, $y$ axes, whilst $c$ is connected with a one-square loop;
\item When a gauge group $G$ is assumed to be continuous one then the
  parameters $a$ and $b$ lead to trivial factors.
\end{itemize}
Therefore, in the first step all (nonequivalent) extensions can be divided
into Abelian ($c=0$) and non-Abelian ($c=1$) ones. Magnetic translation
groups are non-Abelian, hence they can be found amongst groups of the second
type. 

In the above mentioned simplest case $\zz\bigcirc D_2$ all nonequivalent
extensions correspond to all eight types (1 is the unique non-zero element
in $\zz$). As it has been shown in \cite{lch2} there are five letters in the
alphabet $Y$ and they are given by the following letters $A$, $B$, and $C$
introduced in this work
$$
y_1=x_1^2=A_0,\qquad y_2=x_2^2=B_0,\qquad y_3=x_2x_1x_2^{-1}x_1^{-1}=C_{01}
$$ $$ y_4=x_1x_2x_1x_2^{-1}=A_1,\qquad y_5=x_1x_2^2x_1^{-1}=B_1. $$
According with the formulae (\ref{ct}) one obtains
$$ \phi(y_1)=a,\qquad \phi(y_2)=b,\qquad \phi(y_3)=c,\qquad 
  \phi(y_4)=a+c,\qquad \phi(y_5)=b. $$
All possible operator homomorphism, denoted as $\phi_i$ with $i=1,\ldots,8$,
are presented in Table \ref{phi8} rewritten from \cite{lch2} (Table 5) and a
column containing the triple $abc$ is added. 

\begin{table} \caption{\label{phi8} The group of operator homomorphisms 
  $\phi:R\to\zz$} \begin{center} \begin{tabular}{|c|*{6}{|c|}|}\hh
         &$y_1$&$y_2$&$y_3$&$y_4$&$y_5$&$abc$ \nlshh
$\phi_1$ &  0  &  0  &  0  &  0  &  0  & 000\\
$\phi_2$ &  0  &  0  &  1  &  1  &  0  & 001\\
$\phi_3$ &  0  &  1  &  0  &  0  &  1  & 010\\
$\phi_4$ &  0  &  1  &  1  &  1  &  1  & 011\\
$\phi_5$ &  1  &  0  &  0  &  1  &  0  & 100\\
$\phi_6$ &  1  &  0  &  1  &  0  &  0  & 101\\
$\phi_7$ &  1  &  1  &  0  &  1  &  1  & 110\\
$\phi_8$ &  1  &  1  &  1  &  0  &  1  & 111\nlshh
\end{tabular} \end{center} \end{table}

Having this table and using the other results of \cite{lch2} one can easily 
see that homomorphisms with odd indices yield Abelian extensions and that
the first one is simply a direct product $\zz\otimes
D_2\simeq\zz\otimes\zz\otimes\zz \simeq D_{2h}$. The other three are
isomorphic with a direct product $\zz\otimes\zv$, but they differ in orders
of elements $[0,(1,0)]$, $[0,(0,1)]$, and $[0,(1,1)]$. For $\phi_3$
($\phi_5$ and $\phi_7$) order of the first (second and third, respectively)
element is two, whilst it is 4 for the other two elements. For larger
lattices such a classification is a bit more complicated due to many
possible choices of generators for a direct product $\zi1\otimes\zi2$. 

For non-Abelian extensions ($c=1$, \ie\/ $\phi_i$ with even $i$) there are
also two types: (i) $\phi_8$ with $a=b=c=1$ and the extension isomorphic 
with quaternion group (or double dihedral group $D'_2$, \cf\/
\cite{alt1,alt2}) and (ii) extensions isomorphic with the dihedral group
$D_4$. Within the second type one deals with the same isomorphism caused by
an arbitrary choice of generators for $\zz\otimes\zz$: $\{(1,0),(0,1)\}$,
$\{(1,0),(1,1)\}$, and $\{(0,1),(1,1)\}$ for $\phi_2$, $\phi_4$, and
$\phi_6$, respectively. There are at least two facts indicating that this
type of extensions corresponds to a magnetic translation group:
\begin{itemize}
\item The results should be the same (or, in a sense, similar) for a
   continuous group, \ie\/ one has to look for the magnetic translation
   group in types containing extensions with $a=b=0$ and $c\neq0$.  
\item It was shown in Sec \ref{secsch} that unitary operators $U$ and $V$
   have to fulfil condition $U^N=V^N=E$ (see Eq (\ref{epsi})), so --- in the
   considered case --- there should be in $\zz\otimes\zz$ two elements
   $(t_1,t_2)\neq(0,0)$ such that $[0,(t_1,t_2)]^2=[0,(0,0)]$. It is evident
   that there are no such elements in $D'_2$.  
\end{itemize} 
From this follows that the extension with $a=b=0$ and $c=1$. However, one
has to remember that there are many possible choices of generators for
$\zi1\otimes\zi2$ and of $\zn$ what leads to a class of isomorphic
extensions. 

\section{Final Remarks}
It has been shown that the Mac Lane method enables determination of all
nonequivalent extensions of $T=\zi1\times\zi2$ by an Abelian (gauge) group 
$G$. Factor systems $m:T\times T\to G$ can be presented in an analytic form
and they are parametrised by three elements $a,b,c\in G$. It is easy to
notice that some of obtained extensions are isomorphic and this isomorphism
is connected with an arbitrary choice of a generator of the cyclic group
$C_N\subset\uo$. This isomorphism directly corresponds to labelling of basis
vectors in a finite dimensional space $L$ (\cf\/ Sec \ref{secsch}) --- $\ve$
may be not only the primitive root of $1$ but also any power $\ve^k$ of the
primitive root with $\gcd(k,N)=1$.  Moreover, the physical relevance
indicates that extensions by $\uo$ with $k$ not mutually prime with $M$
(\cf\/ Table \ref{ext4}b) should not be taken into account. This problem is
discussed in more details by Wa{\l}cerz \cite{wal} (see also references
quoted therein). The main point is that the factor $\ve$ in Eq (\ref{epsi})
has to be a generator of $C_N$. On the other hand, each linear
representation of a magnetic translation group restricted to the factor
subgroup $G$ should be a faithful one. 

From the considerations presented in this work there follows that the 
parameters $a$, $b$, and $c$ correspond to full loops along the $x$ and $y$ 
axes and to a one `plaquette' loop, respectively. The first two loops lead to 
the zero factors if a continuous group (\eg\/ $\uo$) is assumed to be a gauge
group, but the third one is always present and is relevant to a magnetic
flux. The parameters $a$, $b$, and $c$ provide a classification scheme of
all nonequivalent extensions. The most important role is played by the
parameter $c$ and the extension with $a=b=0$ and $c=1$ can be chosen as a
representative of a class of isomorphic extensions corresponding to magnetic
translation groups.

This work should be completed by investigations of irreducible
representations of studied extensions, especially of these ones, which
describe magnetic translation groups. However, on the one hand, it is too
cumbersome for a brief presentation and, on the other hand, it can be done
applying the standard methods of induced and projective representations,
which can be found in many monographs (\eg\/ \cite{alt3}). Nevertheless, it
can be noticed at the very first glance that representations with the
physical relevance are obtained when a faithful representation of $G$ is
used in the induction procedure. In the considered case of finite gauge
groups $\zn$ it means that one should use one of the irreducible
representation $\Gamma_l(k)=\exp(2\pi\ii kl/N)$ of $\zn$ with $\gcd(l,N)=1$.
Different choices correspond, again, to different generators of $\zn$. It
shows that for $N_1=N_2=N=2$ the two-dimensional representation $E$ of the
dihedral group $D_4$ should be considered (\cf\/ \cite{wal}).

\section*{Acknowledgement}
This work was realized within the project No {\bf PB 695/2/91} supported by
the Polish State Committee for Scientific Research (KBN) in years 1991--93.

\def\theequation{A.\arabic{equation}} \setcounter{equation}{0}
\section*{Appendix: Relations in the Alphabet $Y$} \label{rel}

The alphabet $X$ of the free group $F$ consists of two letters $x_1$ and
$x_2$ such that $M(x_i)=\tau_i$, where $\tau_i$, $i=1,2$, are generators of
the translation group $T$ (see Sec \ref{trgr}). Therefore, 
\eq M(x_i^{N_i})=(0,0)\qquad {\rm for}\quad i=1,2, \qe
so these two words, {\em viz} 
\eq  A := x_1^{N_1},\qquad B := x_2^{N_2},  \label{AB} \qe
belong to the kernel $\Ker M\equiv R\lhd F$. According to the definition
(\ref{hoM}) each of words
\eq  C_{t_1t_2}:=x_1^{t_1}x_2^{t_2}x_1x_2^{-t_2}x_1^{-t_1-1}\label{Fwor}\qe
belongs to $R$, too. For $t_1>0$ one can introduce words 
\eq D_{t_1t_2}:=\prod_{i=0}^{t_1-1}C_{it_2}.  \qe
It is straightforward matter to show that
\eq D_{t_1t_2}:= x_2^{t_2}x_1^{t_1}x_2^{-t_2}x_1^{-t_1} \qe
and, therefore,
\eq x_2^{t_2}x_1^{t_1}=D_{t_1t_2}x_1^{t_1}x_2^{t_2}. \label{com}\qe

All letters of the alphabet $Y$ ({\em cf}\/ Eqs (\ref{AF})) can be expressed
using two letters (\ref{AB}) and $N_1N_2-1$ (for $t_1=0,1,\ldots,N_1-1$ and
$t_2=1,2,\ldots,N_2$ except for the pair $t_1=N_1-1$, $t_2=N_2$) letters $C$
given by Eq (\ref{Fwor}). One can easy check that
\eqa  A_{t_2} &=& C_{N_1-1,t2}A, \\[5pt]
    B_{t_1} &=& D_{t_1N_2}^{-1}B. \aqe
From the above formulae and Eqs (\ref{AF}) follows that
$$  \arr{rclcrcl}
A_0 &=& A, &\qquad& A_{N_2} &=& B_{N_1-1}AB^{-1}, \\[5pt]
B_0 &=& B, & & B_{N_1} &=& ABA^{-1},  \\[5pt]
C_{t_10} &=& 1_F, & & C_{t_1N_2} &=& B_{t_1}B_{t_1+1}^{-1}.
\rra  $$
For example, these relations yield
$$ B_2 = (C_{0N_2}C_{1N_2})^{-1}B = 
  (x_1^2x_2^{N_2}x_1^{-2}x_2^{-N_2})x_2^{N_2} = x_1^2x_2^{N_2}x_1^{-2}.$$

\end{document}